\documentclass[aps,prb,twocolumn,showpacs,floats]{revtex4}
\usepackage{amsmath}
\usepackage{epsfig}
\usepackage{graphics}
\newcommand{\be}{\begin{equation}}
\newcommand{\ee}{\end{equation}}
\newcommand{\bea}{\begin{eqnarray}}
\newcommand{\eea}{\end{eqnarray}}
\newcommand{\ba}{\begin{array}}
\newcommand{\ea}{\end{array}}
\newcommand{\nn}{\nonumber}
\newcommand{\Del}{\Delta}

\newcommand{\Gam}{\Gamma}

\newcommand{\al}{\alpha}
\newcommand{\sig}{\sigma}

\newcommand{\noi}{\noindent}
\newcommand{\eps}{\epsilon}

\newcommand{\ra}{\rangle}
\newcommand{\la}{\langle}

\begin{document}

\title{\bf Energy Spectrum and Exact Cover in an Extended Quantum Ising Model}

\author{
G.M.M. Wakker, R. Ockhorst and M. Blaauboer
}

\affiliation{Delft University of Technology, Kavli Institute of Nanoscience, 
Department of Quantum Nanoscience, 
Lorentzweg 1, 2628 CJ Delft, The Netherlands
}
\date{\today}

\begin{abstract}
We investigate an extended version of
the quantum Ising model which includes beyond-nearest neighbour interactions and an additional
site-dependent longitudinal magnetic field. Treating the interaction exactly and using perturbation 
theory in the longitudinal field, we calculate the energy spectrum and find that the presence of 
beyond-nearest-neighbour interactions enhances the minimum gap between the ground state and 
the first excited state, irrespective of the nature of decay of these interactions along the chain.
The longitudinal field adds a correction to this gap that is independent of the number of qubits. 
We discuss the application of our model to implementing specific instances of 3-satisfiability
problems (Exact Cover) and make a connection to a chain of flux qubits.
\end{abstract}

\maketitle

\section{Introduction}
One of the main motivations for developing scalable quantum processors is the realization that carefully 
constructed quantum algorithms running on such processors can solve certain problems that cannot 
be solved by classical computers~\cite{shor97}. In practice, however, the implementation of quantum 
algorithms using actual qubits - for example solid-state qubits such as spin qubits~\cite{hans07} 
or superconducting circuits~\cite{wend05} - will be hampered by the presence of decoherence, 
which destroys the interference properties
on which succesful execution of these algorithms relies. In order to try to avoid decoherence effects,
Farhi {\it et al.} proposed in 2001 a method of implementing quantum algorithms which relies on the adiabatic 
theorem~\cite{farh01}. The basic idea behind this method, now commonly known as adiabatic quantum 
computation, is to construct a Hamiltonian $H_{\rm final}$ whose (unknown) ground state encodes 
the solution to the problem to be solved. By initializing the qubit system in the known ground state of 
a well-chosen initial Hamiltonian $H_{\rm initial}$ and letting $H_{\rm initial}$ evolve sufficiently slowly 
into $H_{\rm final}$, e.g.
using
\be 
H(t) = \left( 1 - \frac{t}{T} \right) \, H_{\rm initial} + \frac{t}{T}\, H_{\rm final}, 
\label{eq:firstHam} 
\ee 
the system will end up at $t=T$ in the ground state of $H_{\rm final}$. 
Reading out this state then provides the sought-for solution of the problem. 

Since the original proposal of Farhi {\it et al.}, who numerically investigated the required 
running time of adiabatic evolution towards a system whose ground state encodes the solution of 
Exact Cover 3 (a NP-complete~\cite{NP} problem which belongs to the class of 3-satisfiability problems),
a lot of research has been done on adiabatic quantum computation. The efficiency of 
adiabatic quantum computation has been investigated for well-known spin models such 
as the quantum Ising model and the Heisenberg model~\cite{murg04,amin082} and 
the occupation of the ground state has been predicted to be quite robust 
against decoherence (at sufficiently low temperatures and for weak coupling of the qubit to the 
environment)~\cite{chil01,amin08,lloy08}. The relation between adiabatic quantum 
evolution and quantum phase transitions is an ongoing topic of research~\cite{schu06,amin09}. 
Also, recently the statistics and scaling of energy gaps between the ground state and excited states -
which form the limiting factor for the efficiency of adiabatic quantum computation as well as the role played 
by the choice of $H_{\rm final}$ have been investigated~\cite{alts09,knys10,choi10}.

So far, theoretical proposals for the implementation of adiabatic quantum computing have considered
mostly generic spin models, such as the quantum Ising model~\cite{murg04,lieb61}.
These models by themselves cannot be used to encode the solution to one of the hard NP-complete problems
and also in general do not directly correspond to experimental qubit systems, which are often described 
by more complex versions of these spin models~\cite{maje05}.

In this paper we present a first step towards bridging the gap between well-understood generic spin models
and the more complex spin models required for implementing adiabatic quantum computing protocols. 
Specifically, we consider an extended version of the quantum Ising model, which differs from the standard
quantum Ising model in two ways: it allows not only for nearest-neighbour, but also for next-nearest-neighbour
and beyond-next-nearest-neighbour interactions, and it includes an additional site-dependent 
longitudinal magnetic field. Building on a general exact expression for the energy spectrum of 
this extended quantum Ising model for uniform beyond-nearest-neighbour interactions we include the longitudinal
field using perturbation theory. We analyze
the scaling of the energy gap between the ground state and first excited states as a function of 
the number of beyond-nearest-neighbour interactions $M$ and show that the gap increases with $M$ 
both for interactions which decay linearly as a function of the distance between two qubits along 
the chain and interactions which decay exponentially. We then investigate how our model could be 
used to implement and test particular instances of Exact Cover 3 that are characterized by limited 
distance between the bits in each clause (corresponding to the maximum number of beyond-nearest-neighbours 
taken into account). We estimate that the probability of errors to occur is reasonably low ($<$ 10 \%) 
provided enough neighbours are taken into account and enough clauses are defined. We
also discuss the feasibility and prospects of implementing the extended quantum Ising model using a chain of 
superconducting flux qubits.

The paper is organized as follows. In Sec.~\ref{sec-EC} the problem of Exact Cover 3
is introduced, followed by the presentation of our model in Sec.~\ref{sec-model}. 
Sec.~\ref{sec-calc} contains the main calculations: 
the diagonalization of the quantum Ising model with beyond-nearest-neighbour interactions
(Sec.~\ref{sec-diag}), the modification of the resulting energy spectrum by an additional
longitudinal field (Sec.~\ref{sec-long}), and the scaling behavior of the gap 
(Sec.~\ref{sec-scaling}). We then discuss in Sec.~\ref{sec-rel} 
how our model can be used to test particular instances of Exact Cover 3 and make a connection 
to chains of superconducting flux qubits. Conclusions are presented in Sec.~\ref{sec-conc}.

\section{Exact Cover 3}
\label{sec-EC}
Exact Cover 3 belongs to the class of satisfiability problems that are NP-complete~\cite{NP}. 
The problem is the following: a string of $N$ bits $x_1 \ldots x_N$,
 which take values 0 or 1, has to satisfy $M$ constraints called clauses. Each clause 
applies to three bits, say $x_{\al}$, $x_{\beta}$ and $x_{\gamma}$ with $\al$, $\beta$, 
$\gamma$ $\in \{1, \ldots, N\}$ and is satisfied if and only if one of the bits is 1 and 
the other two are 0:
\be
x_{\al} + x_{\beta} + x_{\gamma} = 1.
\ee
The solution of Exact Cover 3, if it exists, consists of an assignment of the bits which
satisfies all of the $M$ clauses. Of particular interest are instances of Exact Cover with a
unique solution\cite{farh01}.

In the literature two types of $H_{\rm final}$ (see Eq.~(\ref{eq:firstHam})) have been considered 
for Exact Cover problems. One involves three-qubit interactions~\cite{farh01}
and the other two-qubit interactions~\cite{banu06,schu06}. In both cases, $H_{\rm final}$
is constructed by associating each violated clause with a fixed energy penalty using the "cost function"
$\sum_{\rm all\ clauses} (x_{\al} + x_{\beta} + x_{\gamma} - 1)^2$. In case of
two-qubit interactions, $H_{\rm final}$ is obtained by replacing $x_{\al}$ by the Ising variables 
$\sigma^{\al}_x = 1 -2x_{\al} = \pm 1$ and substituting $\sigma^{\al}_x$ by the Pauli operators 
$\hat{\sigma}^{\al}_x$. This yields (omitting an irrelevant constant)~\cite{schu06} 
\be
H_{\rm final} = \frac{1}{4} \sum_{\al,\beta=1}^{N} M_{\al \beta} \sig^{\al}_x \sig^{\beta}_x
- \frac{1}{2} \sum_{\al=1}^{N} N_{\al} \sig^{\al}_x,
\label{eq:Hfinal}
\ee
where $\sig^{\al}_x$ denotes the Pauli matrix for qubit $\al$ (omitting the hat), $N_{\al}$ represents 
the number of clauses involving qubit $\al$ and $M_{\al \beta}$ denotes the number of clauses which involve both
qubit $\al$ and qubit $\beta$.

From an experimental point of view, both the Hamiltonian involving three-qubit interactions from 
Ref.~\cite{farh01} and the Hamiltonian (\ref{eq:Hfinal}) are not easy to realize. In existing solid-state qubit
systems so far three-qubit interactions have not been realized yet. 
Common potentially scalable qubit systems, e.g. electron spin qubits~\cite{hans07} or superconducting 
qubits~\cite{wend05} involve two-qubit interactions whose strength is a function of the 
distance between the qubits rather than dictated by the clauses 
(as in Hamiltonian (\ref{eq:Hfinal})). All in all, theoretical predictions of Exact Cover 3 and 
other 3-satisfiability problems still seem somewhat removed from experimental verification. 
The aim of this paper is to provide a first step towards bridging this gap between theory 
and experiment, by analyzing the model Hamiltonian 
(an extended version of the quantum Ising model) that is introduced in the next section.

\section{Model}
\label{sec-model}

\noi Our starting point is the time-dependent spin-chain Hamiltonian
\be
H(t) = f(t)\, \Delta \sum_{i=1}^N \sigma_z^i + g(t) \left( \sum_{i,j=1}^{N} J_{ij} \sigma_x^i
\sigma_x^j + \sum_{i=1}^N h_i \sigma_x^i \right).
\label{eq:Ham}
\ee
Here $N$ denotes the number of qubits along the chain, $J_{ij}$ represents the Ising interaction 
between qubit $i$ and qubit $j$, $\Del$ denotes a transverse magnetic field and $h_{i}$ is 
a site-dependent longitudinal field. The functions $f(t)$ and $g(t)$ 
model  the time evolution from $t=0$ to $t=T$. In most of this paper we choose $f(t) = C- t/T$,
with $C$ a constant, and $g(t) = t/T$. When we deviate from this time dependence, this is 
indicated in the text. For any $0 \leq t \leq T$ the instantaneous 
Hamiltonian (\ref{eq:Ham}) represents an extended quantum Ising model which includes
a site-dependent longitudinal field $h_i$ and whose interaction term not only allows for 
nearest-neighbour interaction $J_{i,i+1}$ but also for next-nearest-neighbour interactions
and beyond.
A similar model was recently considered by Amin and Choi~\cite{amin09}, who investigated the 
occurrence of first order quantum phase transitions in an inhomogeneous version of the 
Hamiltonian~(\ref{eq:Ham}). The standard quantum Ising model $H_{\rm Ising}= 
J \sum_{i=1}^{N} \sigma_z^i \sigma_z^{i+1} + \Del
 \sum_{i=1}^{N} \sigma_x^i$  with fixed nearest-neighbour interactions is a well-known and
exactly solvable spin model~\cite{lieb61} which has been studied extensively for more than 
50 years. In the context of adiabatic quantum computing, Murg and Cirac~\cite{murg04} have investigated 
adiabatic evolution in the quantum Ising model using the ratio $\Delta/J$ as the time-dependent 
parameter and calculated the excitation probability from the ground state to higher-energy 
states. More recently the robustness of adiabatic passage against noise was studied~\cite{fubi07}.

Eq.~(\ref{eq:Ham}) represents a chain of qubits which initially at $t=0$ is described by the 
Hamiltonian
\be
H_{\rm initial} \equiv \tilde{\Gam} \sum_{i=1}^N \sigma_z^i
\label{eq:initial}
\ee
and has evolved after time $t=T$ into the Hamiltonian
\be
H_{\rm final} \equiv \Gam \sum_{i=1}^N \sigma_z^i + \sum_{i,j=1}^N J_{ij} 
\sigma_x^i \sigma_x^j + \sum_{i=1}^N h_i \sigma_x^i,
\label{eq:finalHamiltonian}
\ee
with $\tilde{\Gam} \equiv C\Delta$ and $\Gam \equiv (C-1) \Delta$. 
The initial Hamiltonian Eq.~(\ref{eq:initial}) describes a chain of spins in a magnetic field
directed along the $z$-axis, whose ferromagnetic ground state 
consists of a large superposition of states. The final Hamiltonian Eq.~(\ref{eq:finalHamiltonian})
reduces to the Hamiltonian (\ref{eq:Hfinal}) for $\Gam =0$ (i.e. $C=1$) and 
thus encodes the solution of a particular instance of Exact Cover 3 if $h_i$ ($J_{ij}$)
is interpreted as the number of clauses containing bit $i$ (both bit $i$ and bit $j$).
For $J_{ij} \equiv J$ and $h_i \equiv h$ $\forall i,j$ site-independent, the Hamiltonian 
Eq.~(\ref{eq:finalHamiltonian}) reduces to a quantum Ising model in an additional uniform 
longitudinal field $h$. 
Using perturbation theory in $h$, the ground state of this Hamiltonian and the 
scaling behavior of the gap has been studied in Ref.~\cite{ovch03}.

\section{Calculations}
\label{sec-calc}

In this section we first diagonalize the Hamiltonian (\ref{eq:finalHamiltonian}) in absence 
of the longitudinal field $h_i$, assuming uniform nearest-neighbour interactions and uniform 
next-nearest-neighbour three-qubit interactions~\cite{threequbit}. From the energy spectrum 
we calculate the energy gap between the ground state and the first exited state and derive 
the condition for this gap to be minimal. In Sec.~\ref{sec-long} we then include the site-independent 
longitudinal field $h_i$ and calculate the corrections to the energy spectrum due to this field up 
to second order in perturbation theory. In Sec. \ref{sec-scaling} we use this modified spectrum 
to analyze the scaling behavior of the gap as a function of the coupling strengths $\lambda_j$.

\subsection{Diagonalization}
\label{sec-diag}

\noi Our starting point is the Hamiltonian:
\be
H_0  =  \Gamma \sum_{i=1}^N \sig_z^{i}
+ \sum_{i=1}^N \left( J_1\, \sig_x^{i} \sig_x^{i+1}
+ J_2\, \sig_x^{i} \sig_z^{i+1} \sig_x^{i+2} \right). 
\label{eq:H_threequbit}
\ee
$H_0$ originates from the Hamiltonian $H_{\rm final}$ [Eq.~(\ref{eq:finalHamiltonian})] by taking 
$h_i=0 \; \; \forall i$, defining 
$J_1 \equiv J_{i,i+1}$, $J_2 \equiv J_{i,i+2} \; \; \forall i$, taking $J_{i,j}=0$ otherwise and
adding the third-qubit interaction in the $J_2$-term. By applying a Jordan-Wigner transformation 
to $H_0$, Eq.~(\ref{eq:H_threequbit}) can be rewritten in bilinear form as (omitting an overall 
minus-sign)~\cite{suzu71}
\bea
H_0 & = & \frac{\Gamma N}{2} - \Gamma \sum_{i=1}^N c_i^{\dagger} c_i -\frac{J_1}{4} \sum_{i=1}^N 
(c_i^{\dagger} - c_i)(c_{i+1}^{\dagger} + c_{i+1}) \nn \\
& + & \frac{J_2}{4} \sum_{i=1}^N (c_i^{\dagger} - c_i)(c_{i+2}^{\dagger} + c_{i+2}) \nn \\
& + & \frac{J_1}{4} (c_N^{\dagger} - c_N)(c_1^{\dagger} + c_1)\, (e^{i\pi L} + 1) \nn \\
& - & \frac{J_2}{4} (c_N^{\dagger} - c_N)(c_2^{\dagger} + c_2)\, (e^{i\pi L} + 1) \nn \\
& - & \frac{J_2}{4} (c_{N-1}^{\dagger} - c_{N-1})(c_1^{\dagger} + c_1)\, (e^{i\pi L} + 1).
\label{Ham_neighbours}
\eea
Here $c_i^{\dagger}$ and $c_i$ denote fermionic raising and lowering operators and 
$L \equiv \sum_{j=1}^N c_j^{\dagger} c_j$ as in Ref.~\cite{pfeu70}. 
The last three terms are absent in case of periodic boundary conditions, and 
can be neglected for $N \gg 1$. Diagonalizing (\ref{Ham_neighbours}) using Pfeuty's method~\cite{pfeu70} yields: 
\be
\frac{H_0}{\Gamma} = \sum_k \Lambda_k \eta_k^{\dagger} \eta_k - \frac{1}{2} \sum_k \Lambda_k,
\label{Diag12}
\ee
with the fermionic operators
\be
\eta_k = \sum_{i=1}^N \left\{ \left( \frac{\phi_{ki} + \psi_{ki}}{2} \right) c_i + 
\left( \frac{\phi_{ki} - \psi_{ki}}{2} \right) c_i^{\dagger} \right\}.
\label{eta_k}  
\ee
For $N$ even, the sums over $k$ in Eq.~(\ref{Diag12}) run from $-N/2$ to $(N-2)/2$. 
For $N$ odd  the sums run from $(1-N)/2$ to $(N-1)/2$. Defining $\lambda_j \equiv J_j/(2\Gamma)$,
the functions $\phi_{ki}$ and $\psi_{ki}$ are given by
\begin{subequations}
\bea
\phi_{ki} &=& \begin{cases} \sqrt{\frac{2}{N}} \sin \left(\frac{2 \pi i k}{N} \right) \; \; \; \; \; \textrm{for} \; \; k>0 \\
\sqrt{\frac{2}{N}} \cos \left(\frac{2 \pi i k}{N} \right) \; \; \; \; \; \textrm{for} \; \; k\leq0 \end{cases} \\
\nonumber \psi_{ki} &=& -\frac{1}{\Lambda_k} \left\{ \left[ 1 +\lambda_1 \cos\left( \frac{2 \pi k}{N} \right) -
\lambda_2 \cos\left( \frac{4 \pi k}{N} \right) \right] \phi_{ki} \right. \\
& & + \left. \left[ \lambda_1 \sin\left( \frac{2 \pi k}{N} \right) -
\lambda_2 \sin\left( \frac{4 \pi k}{N} \right) \right] \phi_{(-k)i} \right\} 
\eea
\label{defin}
\end{subequations}
and the energy eigenvalues are
\bea
\nonumber \Lambda_k^2 &=& \left[ 1 +\lambda_1 \cos\left( \frac{2 \pi k}{N} \right) -
\lambda_2 \cos\left( \frac{4 \pi k}{N} \right) \right]^2 + \\
\nonumber & & \left[ \lambda_1 \sin\left( \frac{2 \pi k}{N} \right) -
\lambda_2 \sin\left( \frac{4 \pi k}{N} \right) \right]^2 \\
\nonumber &=& 1+ \lambda_1^2 + \lambda_2^2 + 2 \lambda_1(1-\lambda_2) \cos\left( \frac{2 \pi k}{N} \right) \\
& & -2\lambda_2 \cos\left( \frac{4 \pi k}{N} \right).
\label{eq:Diagham_neighbours}
\eea
The diagonalization procedure leading to the energy spectrum (\ref{eq:Diagham_neighbours}) can be 
generalized for higher $\lambda_j$, $j=3,4,\ldots$ (i.e. including interactions between qubits that 
are farther apart) by adding terms 
\be
\nonumber (-1)^{j+1} \lambda_j \cos\left( \frac{2 \pi j  k}{N} \right) \; \; \textrm{and} \; \; 
(-1)^{j+1} \lambda_j \sin\left( \frac{2 \pi j  k}{N} \right)   
\ee
inside the square brackets in the expressions for $\psi_{ki}$ and $\Lambda_k^2$. However, 
one should keep the number $M$ of neighbours included much smaller than $N/2$, in order to be 
able to neglect the boundary terms in Eq.~(\ref{Ham_neighbours}). The full expression for the 
energy spectrum now reads~\cite{suzu71}
\bea
\nonumber \Lambda_k^2 &=& \left[ 1 + \sum_{j=1}^{M} (-1)^{j+1} \lambda_j \cos\left( 
\frac{2 \pi j  k}{N} \right) \right]^2 + \\
& & \left[ \sum_{j=1}^{M} (-1)^{j+1} \lambda_j \sin\left( \frac{2 \pi j  k}{N} \right) \right]^2.
\label{eq:Diagham_fullneighbours}
\eea
Note that when $N$ is even, $k=-N/2$, and $\sum_{j=1}^{M} \lambda_j =1$  we obtain from 
Eq.~(\ref{eq:Diagham_fullneighbours}) 
that $\Lambda_k=0$. In that case the groundstate and first excited state are degenerate and the 
concept of adiabatic transport no longer applies~\cite{pfeu702}.  
After the inclusion of the longitudinal field (see Eq.~(\ref{eq:finalHamiltonian})) term
in the Hamiltonian $H_0$ this degeneracy is lifted, as we show below in Section~\ref{sec-long}. For now, however, 
we restrict ourselves to the case $N$ odd. From Eq.~(\ref{Diag12}) we obtain that the ground state 
energy of the system is given by
\be
E_g = -\frac{\Gamma}{2}\, \sum_k \Lambda_k.
\ee
The energy difference between the ground state and the excited single $k$-fermion state is $\Lambda_k$ for 
$k$ running from $-(N-1)/2$ to $(N-1)/2$. The minimum gap $\Del_{\rm eg,min}$ between the ground state and the
first excited state is then derived by minimizing $\Lambda_k$ with respect to a continuous variable $k$. 
Including up to next-nearest-neighbour interactions, the minimum is attained at $k=\pm(N-1)/2$ for 
$\lambda_2 < 1$ and at $k=0$ for $\lambda_2 > 1$. In what follows, we proceed with the case 
$\lambda_2 <1$, which corresponds to the most 
physical situation. Including nearest-neighbours only, the energy gap $\Delta_{\rm eg}$ is then given by: 
\bea
\nonumber \Delta_{\rm eg} \! \! \! \! \! &=& \! \! \! \!\! \Gamma \sqrt{1+ \lambda_1^2 + \lambda_2^2 - 
2 \lambda_1(1-\lambda_2) \cos\!\left(\! \frac{\pi}{N} \! \right) - 2 \lambda_2 \cos\!\left(\! \frac{2\pi}{N} \! \right)}, \\
&\stackrel{N \gg 1}{\approx}& \! \!  \Gamma \sqrt{ (\lambda_1 + \lambda_2 -1)^2 +(\lambda_1 - 
\lambda_1 \lambda_2 + 4 \lambda_2)\left( \frac{\pi}{N} \right)^2}.
\label{gapm}
\eea 
This gap is minimal for $\lambda_1 + \lambda_2 =1$, and then equals 
\be
\Del_{\rm eg,min} \! = \! \frac{\pi \Gamma}{N}\, \sqrt{\lambda_1 -\lambda_1 \lambda_2 + 4 \lambda_2}  = 
\! \frac{\pi \Gamma}{N}\, (1+ \lambda_2).
\label{gapmin}
\ee
For coupling with more neighbours while keeping $N \gg M$, Eq.~(\ref{gapmin}) generalizes to
\be
\Del_{\rm eg,min}(M) = \frac{\pi \Gamma}{N}\, \sum_{j=1}^{M} j \lambda_j, \ \  \rm with\ 
\sum_{j=1}^M \lambda_j =1.
\label{gap}
\ee
\begin{figure}
\begin{center}
\centering
\scalebox{0.95}{\includegraphics{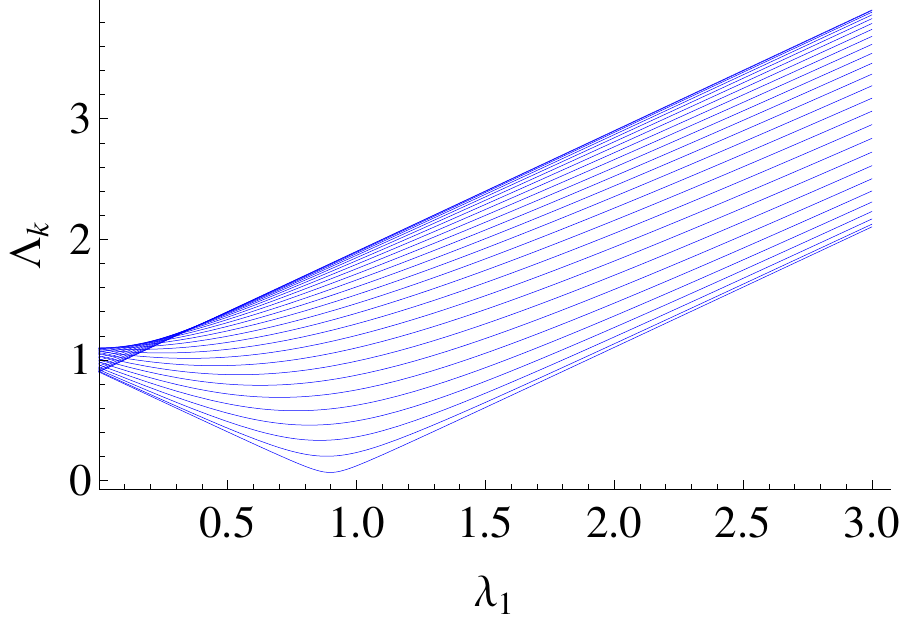}} 
\caption{(color online) Plot of the single-fermion energy levels $\Lambda_k$ [Eq.~(\ref{eq:Diagham_neighbours})] as a 
function of $\lambda_1$ for $N=51$, $\lambda_2 =0.1$ and $k$ ranging from 0 to $\pm$ 25. 
The lowest graph represents $k=\pm25$. From top to bottom (on the right hand side of the plot) the graphs are ordered 
from lowest ($k=0$) to highest ranging ($k=\pm25$) values.} 
\end{center}
\end{figure}
The minimum gap $\Del_{\rm eg,min}(M)$ is thus inversely proportional to the number of qubits~\cite{murg2},
irrespective of the number of nearest-neighbour interactions $M$. From Eq.~(\ref{gap}) we obtain that 
adding beyond-nearest-neighbour interactions increases the minimum gap by
\bea
\Delta_{\rm eg,min}(M) - \Delta_{\rm eg,min}(M-1) & = & \frac{\pi \Gamma}{N}\, \left( 1 - \sum_{j=1}^{M-1} \lambda_j \right) 
\nn \\ & = & \frac{\pi \Gamma \lambda_M}{N} > 0.
\eea
In Fig. 1 the energy levels for the single-fermion excited states (\ref{eq:Diagham_neighbours}) are plotted as a function 
of $\lambda_1$ for fixed $\lambda_2$. The minimum gap indeed occurs for $\lambda_1 + \lambda_2 =1$.

\subsection{Perturbation theory in the longitudinal field}
\label{sec-long}

We now apply perturbation theory to calculate the effect of the longitudinal field $h_i$ [the third term in 
Eq.~(\ref{eq:finalHamiltonian}) which we denote by $H_3$] on the level spectrum (\ref{eq:Diagham_fullneighbours}). 
In terms of the fermionic operators $\eta_k^{\dagger}$ and $\eta_k$ in Eq.~(\ref{eta_k}), $H_3$ is written as:
\bea
\nonumber \frac{H_3}{\Gamma} &=& -\sum_{i=1}^N \frac{h_{i}}{2\Gamma} (-1)^{N_f} \left( c_i^{\dagger} + c_i \right) \\
  &=& - \sum_k r_k \left( \eta_k^{\dagger} + \eta_k \right),
\label{eq:minus1}
\eea
with 
\be 
r_k \equiv \sum_{i=1}^N \frac{h_{i}}{2\Gamma} (-1)^{N_f} \phi^{-1}_{ik},
\label{rparameter}
\ee
$N_f$ the total number of $c$-fermions and $\phi^{-1}_{ik}$ given by Eq.~(\ref{eq:inverse_phi_psi}). 
We denote the vacuum state by $| 0 \rangle$ and the state with one $\eta$-fermion 
by $|\theta_{p}\rangle$. For more $\eta$ fermions we use more indices, 
for example $|\theta_{p,q,r}\rangle$. To first order in $h_{i}$, the correction of the energy of the vacuum state is given by
\be
\delta E_0^{(1)} = \left\la 0 \left|H_3\right| 0 \right\ra \; \; = \; \; 0.
\ee
Analogously, the first-order correction of the energy corresponding to 
state $|\theta_m \ra$ is given by~\cite{ovch03}:
\bea
\delta E_m^{(1)} &=& \left\la \theta_m \left|H_3\right| \theta_m \right\ra \nonumber \\
\nonumber &=& -\Gamma \la \theta_m | \sum_k r_k \left( \eta_k^{\dagger} + \eta_k \right)| \theta_m \ra \\
\nonumber &=& -\Gamma \left( \sum_{k \neq m} r_k \la \theta_m | \theta_{k,m} \ra + r_m \la \theta_m |0 \ra \right) \\
& = & 0. \nonumber
\label{firstcor}
\eea
This line of reasoning can be extended to all odd-order corrections of the energy levels, see Appendix~\ref{appendix}. 
The lowest nonzero correction to the energy spectrum is thus the second-order correction. For the ground state 
$|0\ra$ this is given by:
\bea
\nonumber \delta E_0^{(2)} &=&  \Gamma^2 \sum_{k} \frac{\bigg\vert \la 0 | \sum_{l} r_l \left( \eta_l^{\dagger} + 
\eta_l \right) | \theta_k \ra \bigg\vert^2}{E_0 - E_k} \\
&=& - \Gamma \sum_{k}\frac{r_k^2}{\Lambda_k}.
\label{02corr}
\eea
To second order the corrections to the energies of the single fermion states $|\theta_m\ra$ are given by:
\bea
\nonumber \delta E_m^{(2)} &=& \Gamma^2 \left[ \frac{|\la \theta_m |r_m \eta_m^{\dagger} | 0 \ra |^2}{E_m-E_0} 
+ \sum_{k \neq m} \frac{| \la \theta_m|r_k \eta_k | \theta_{m,k} \ra|^2}{E_m-E_{mk}} \right]\\
&=& \Gamma \left[ 2 \frac{r_m^2}{\Lambda_m} - \sum_{k} \left( \frac{r_k^2}{\Lambda_k} \right) \right].
\label{02mcorr}
\eea
Higher-order corrections and a discussion of the validity of perturbation theory are given in Appendix~\ref{appendix}. 
For a site-independent longitudinal field $h_i \equiv h \; \forall i$ we obtain from Eq.~(\ref{rparameter}): 
\bea
r_k &=& \begin{cases} \ba{lll} 
\frac{h}{2 \Gamma} \sqrt{\frac{2}{N}} \tan \left( \frac{k \pi}{N} \right) & \textrm{for} & k>0 \\  
\frac{h}{4 \Gamma} \sqrt{\frac{2}{N}}  &  \textrm{for} &  k=0 \\ 
\frac{h}{2 \Gamma} \sqrt{\frac{2}{N}}  & \textrm{for} & k<0 \ea \end{cases}
\label{rk}
\eea
Here we have used the inverse of the functions $\phi_{ki}$ and $\psi_{ki}$ [Eqns.~(\ref{defin})], that are defined by  
\be
\sum_{k} \phi^{-1}_{ik} \phi_{ki} = \sum_{k} \psi^{-1}_{ik} \psi_{ki} = 1 \; \; \forall i,
\ee 
and are given by (for arbitrary $N$)
\bea
\nonumber \phi^{-1}_{ik} &=& \left( 1 - \frac{1}{2} \delta_{0k} - \frac{1}{2} \delta_{\frac{N}{2} k}\right) \phi_{ki}  \\
\nonumber \psi^{-1}_{ik} &=& - \left[ 1 +\lambda_1 \cos\left( \frac{2 \pi k}{N} \right) -
\lambda_2 \cos\left( \frac{4 \pi k}{N} \right) \right] \frac{\phi^{-1}_{ik}}{\Lambda_k} \\
&&- \left[ \lambda_1 \sin\left( \frac{2 \pi k}{N} \right) -
\lambda_2 \sin\left( \frac{4 \pi k}{N} \right) \right] \frac{\phi^{-1}_{i(-k)}}{\Lambda_k} \; \; \; 
\label{eq:inverse_phi_psi}
\eea
Finally, we note that Eq.~(\ref{rparameter}) can also be used to calculate $r_k$ if the longitudinal field is site-dependent.
For example, if $h_i$ has a given statistical distribution, Eq.~(\ref{rparameter}) can be used to calculate the corresponding 
distribution and average of $r_k$.

\subsection{Scaling behavior of the gap}
\label{sec-scaling}

Using Eqns.~(\ref{02corr}) and (\ref{02mcorr}) we now investigate the second-order correction of the gap 
$\delta \Delta_{eg,m}^{(2)}$ 
between the ground state and the $m^{th}$ single-fermion state:
\bea
\delta \Delta_{{\rm eg},m}^{(2)} & \equiv & \delta E_{m}^{(2)} - \delta E_0^{(2)} \nn \\
& = & 2\, \Gamma \frac{r_m^2}{\Lambda_m} \nn \\
& = & \frac{h^2}{\Gamma N} \left\{ \ba{ll}
\frac{\tan^2 \left( \frac{m \pi}{N} \right)}{\Lambda_m} & m > 0 \\
\frac{1}{4 \Lambda_0} & m=0. \\
\frac{1}{\Lambda_m} & m < 0
\ea \right. 
\label{eq:deltaDelta}
\eea
From Sec.~\ref{sec-diag} we know that the minimum gap in the absence of the longitudinal field occurs for 
$m = \pm \frac{N-1}{2}$ and $\sum_{j=1}^M \lambda_j=1$. The longitudinal field lifts the degenracy of 
the $\pm (\frac{N-1}{2}$)-fermion levels. Using Eq.~(\ref{eq:deltaDelta}) we then obtain for the corresponding 
corrections of the gap (at $\sum_{j=1}^M \lambda_j=1$):
\begin{subequations}
\bea
\delta \Delta_{{\rm eg},\frac{N-1}{2}}^{(2)} & = & \frac{h^2}{\Gamma N}\, \frac{\tan^2 \left( \frac{\pi}{2} - 
\frac{\pi}{2N} \right)}
{\Lambda_{\frac{N-1}{2}}} \nn \\
& \approx & \frac{4 h^2 N}{\pi \Gamma}\, \frac{1}{\Lambda_{\frac{N-1}{2}}} \\
\delta \Delta_{{\rm eg},-\frac{N-1}{2}}^{(2)} & = & \frac{h^2}{\Gamma N}\, \frac{1}
{\Lambda_{\frac{N-1}{2}}} <  \delta \Delta_{{\rm eg},\frac{N-1}{2}}^{(2)}. \nn \\
\label{corrmingap}
\eea
\end{subequations}
Substituting Eq.~(\ref{eq:Diagham_fullneighbours}) for $k=\frac{N-1}{2}$ into Eq.~(\ref{corrmingap}), 
the second-order correction to the minimum gap is  thus given by:
\be
\delta \Delta_{\rm eg,min}^{(2)}(M) = \frac{h^2}{\pi \Gamma}\, \frac{1}{\sum_{j=1}^M j \lambda_j}.
\label{minimumgap2}
\ee
Combining Eqns.~(\ref{gap}) and (\ref{minimumgap2}) we then obtain for the minimum gap, up to second 
order in the longitudinal field $h$,
\be
\Delta_{\rm eg,min}(M) = \frac{\pi \Gamma}{N} \sum_{j=1}^M j \lambda_j + \frac{h^2}{\pi \Gamma}\, 
\frac{1}{\sum_{j=1}^M j \lambda_j},
\label{minimumgap}
\ee
with $\sum_{j=1}^M \lambda_j = 1$. Eq.~(\ref{minimumgap}) is the main result of our paper. We see that 
the presence of the longitudinal field leads to an increase of the gap by a factor that is independent of the 
number of qubits $N$. Although the $h$-dependent correction term Eq.~(\ref{minimumgap2}) decreases 
when adding beyond-nearest-neighbour interactions (scaling as $\delta \Delta^{(2)}_{\rm eg,min} (M)- 
\delta \Delta^{(2)}_{\rm eg,min}(M-1) \sim - h^2 \lambda_M/(\pi \Gamma) < 0$) the minimum gap 
itself increases with $M$ since the first, unperturbed, term increases with $\lambda_M$.
\begin{figure}
\begin{center}
\centering
\scalebox{0.95}{\includegraphics{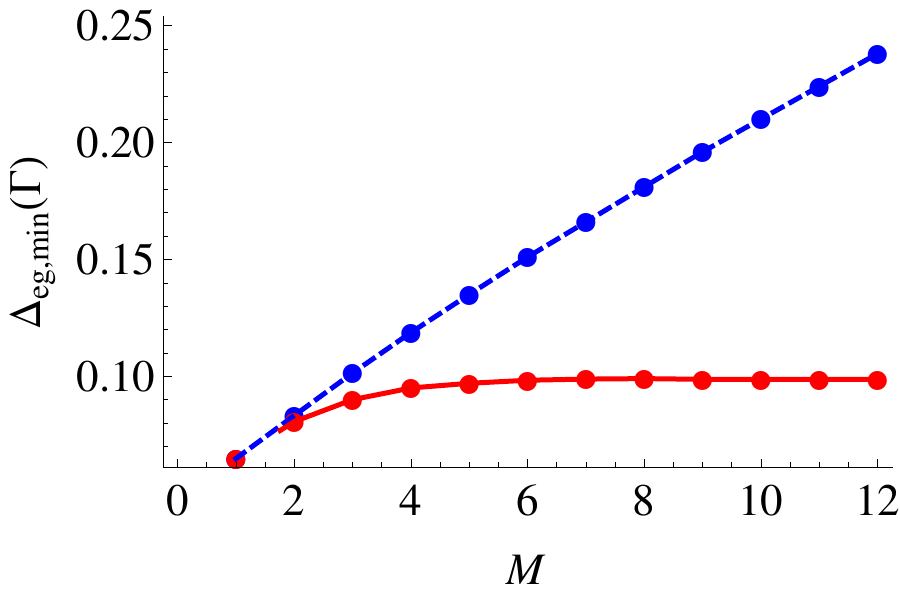}} 
\caption{(color online) Plot of the minimum gap $\Delta_{\rm eg,min}$ [Eq.~(\ref{minimumgap})] in units of 
$\Gamma$ as a function of $M$ for $h/\Gamma = 0.1$ and $N=51$. The blue dashed line represents linear 
($\lambda_j = \textrm{const}/j$) and the red solid line represents exponential 
($\lambda_j = \textrm{const} \cdot \exp{(-j)}$) decay of the interaction strength with distance between the bits. 
The constant is determined by the condition $\sum_{j=1}^M \lambda_j=1$.
} 
\end{center}
\label{fig:mingap}
\end{figure}
Fig.~\ref{fig:mingap} depicts $\Delta_{eg,min}$ in units of $\Gamma$ as a function of the number of neighbouring 
interactions $M$. We consider both interactions which decay linearly as a function of the distance between two qubits 
along the chain and interactions which decay exponentially. For both types of decay (although more strongly for linear 
than exponential decay) the minimum gap is enhanced by including coupling with more neighbours. Since the required 
running time of an adiabatic algorithm is inversely proportional to the square of the energy gap~\cite{murg04}, 
the enhancement of the gap implies that a qubit system with beyond-nearest-neighbour interactions may be 
advantageous for implementing adiabatic quantum algorithms.

\section{Relation to Exact Cover 3}
\label{sec-rel}

In this section we investigate how the extended quantum Ising model from the previous section with uniform 
beyond-nearest-neighbour interactions can be used to simulate specific instances of Exact Cover 3. In particular, 
we numerically estimate the probability for obtaining the correct solution of instances of Exact Cover 3 that are 
characterized by a maximum distance between bits in a clause as a function of the number $M$ of 
beyond-nearest-neighbour interactions that are included. At the end of the section we make a connection 
to an actual experimental system (a chain of flux qubits).

We consider a system of $N$ bits with coupling to $M$ nearest neighbours. Since each bit is coupled to $M$ bits on either side, 
the maximum distance between two bits that appear in the same clause is $2M$ (see also Eq.~(\ref{eq:Hfinal})). 
Out of the set of all possible clauses, we restrict ourselves to clauses that satisfy this property, i.e. contain bits which 
are at most a distance $2M$ along the chain apart. We refer to this subset as the set of "restricted clauses". 
Clauses in which no maximum distance between the bits is defined, are called "unrestricted clauses". 
For clauses $\{ \alpha , \beta , \gamma \}$ with $\alpha < \beta < \gamma$ we only allow restricted clauses 
that satisfy $\gamma-\alpha \leq 2M$ or $(\alpha+N)-\beta \leq 2M$ or $(\beta+N)-\gamma \leq 2M$ and 
in order to avoid boundary effects we consider a cyclic chain of bits. The question that we raise is the following: 
what is the probability that a particular solution of Exact Cover 3, represented by a random bit chain (in which 
each bit independently has a probability $p$ for taking value $1$) can be reconstructed using $K$ restricted clauses? 
In our simulations the restricted clauses for given $K$ are selected from a large group of unrestricted clauses
that are uniformly distributed over the $N$ bits in the qubit chain~\cite{meza02}. For some limits the answer 
to this question is obvious: in particular, for $p$ close to $0$ or $1$ there are too many bits with the same value, 
and no restricted set of clauses, or indeed unrestricted set of clauses, can be found that gives a uniquely satisfying assignment.

We now describe our simulations. For specific values of $N$, $M$, $p$, and $K$ we generate $100$ random 
bit strings~\cite{bitstrings}. For each of these we randomly generate $500$ clauses (or more if needed) that are satisfied by
the given bit string. Out of these we randomly select $K$ clauses that satisfy the 
$M$-nearest-neighbour restriction condition. Then we check whether the original bit string is the unique solution of 
the restricted set of clauses. If each bit is covered by at least one clause, the solution is usually unique. 
Out of $100$ runs we deduce a probability $p_E$ that an error will occur, meaning that at least one other 
bit string also satisfies the assignment of $K$ clauses. Our goal is to determine at which point $p_E$ makes 
a transition from large ($p_E \approx 1$) to small ($p_E \approx 0$) as a function of the parameters $p$ and $M$. 
In the table below, we keep $N$ and $K$ fixed at $N=12$ and $K=20$, and calculate $p_E$ as a function of $K$ 
for various probabilities $p$. The ratio $K/N$ is chosen large enough that once the bits allow a uniquely satisfying 
assignment of restricted clauses for a given $p$, this is also formed in the simulations with high probability.

\begin{table}
\begin{center}
  \begin{tabular}{ | c || c | c |c | c |c | c |c | c | }
    \hline
    $M$ & 1 & 2 & 3 & 4 & 5 & 6 & 7 & 8 \\ \hline
     $p_E$ \rm for $p=0.2$ & 1.00 & 0.64 & 0.36 & 0.22 & 0.14 & 0.14 & 0.10 & 0.09 \\ \hline 
      $p_E$ \rm for $p=0.3$ & 1.00 & 0.43 & 0.22 & 0.11 & 0.10 & 0.10 & 0.09 & 0.05 \\ \hline 
       $p_E$ \rm for $p=0.4$ & 1.00 & 0.51 & 0.26 & 0.18 & 0.16 & 0.10 & 0.10 & 0.09 \\ \hline 
        $p_E$ \rm for $p=0.5$ & 1.00 & 0.67 & 0.34 & 0.23 & 0.22 & 0.18 & 0.17 & 0.13 \\ 
    \hline
    \end{tabular}
\end{center}
\caption{The probability $p_E$ for errors to occur as a function of the number $M$ of beyond-nearest-neighbour 
interactions for $p=0.2, 0.3, 0.4,$ and $0.5$.} 
\label{firsttable}
\end{table}
\begin{figure}
\begin{center}
\centering
\scalebox{0.95}{\includegraphics{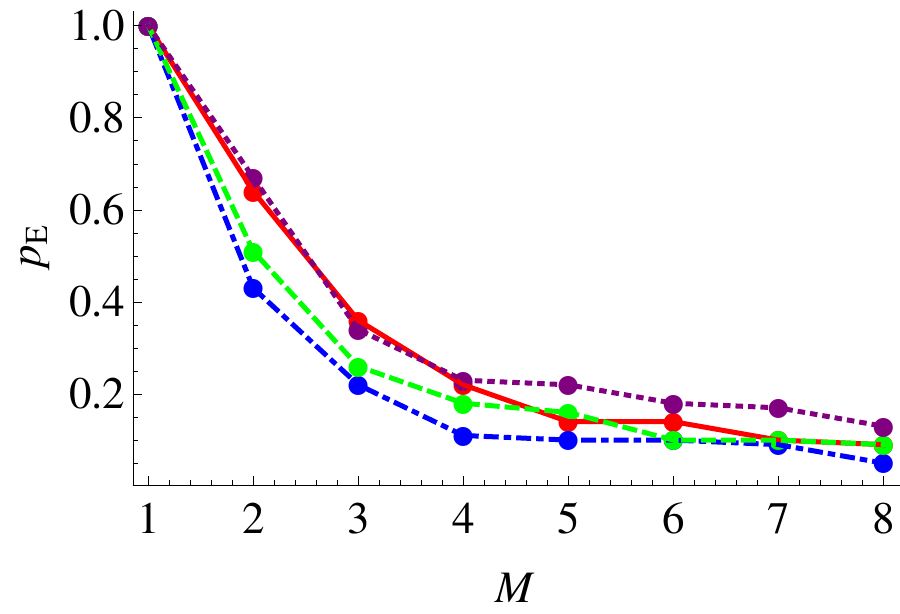}} 
\caption{(color online) The probability $p_E$ from Table~\ref{firsttable} as a function of $M$ 
for $p=0.2$ (red, solid), $p=0.3$ (blue, dot-dashed), $p=0.4$ (green, dashed), and $p=0.5$ (purple, dotted).
} 
\end{center}
\label{fig:probability}
\end{figure}
From Table~\ref{firsttable} we see that for all values of $p$ a transition from large to small error probability 
takes places for $M$ ranging from $1$ to $8$. For $p=0.3$ the probability of errors to occur for a given value 
of $M$ is smallest, which can be explained from the optimal ratio of bitvalues $0$ and $1$ for finding clauses. 
Fig.~\ref{fig:probability} shows a graphic representation of Table~\ref{firsttable}. We see that the probability 
for an error to occur during simulation of restricted  instances of Exact Cover 3 decreases approximately 
exponentially with increasing $M$.

Our simulations did not take into account the fact that the interaction strength $\lambda_n$ between qubits 
$i$ and $i+n$ - which translates into the number of clauses that contain both bit $i$ and bit $i+n$, 
see Eq.~(\ref{eq:Hfinal}) - in practice decreases as a function of the distance $n$ between the qubits. 
In order to make a connection to 
the Exact Cover Hamiltonian~(\ref{eq:Hfinal}) we in principle thus need to further restrict the sets of "restricted clauses" 
to sets in which the number of clauses containing nearby qubits is larger than the number of clauses containing bits 
that are farther apart. Although we did not investigate this in depth, back-of-the-envelope estimates indicate that 
in order to reach the same success probability a factor of $2$ more clauses need to be included in the simulations.

In order to simulate Exact Cover 3 in practice, one needs to use adiabatic evolution of the Hamiltonian 
(\ref{eq:Ham}), whose final ground state at time $T$ encodes the solution of the Exact Cover problem. 
In their general form, the Hamiltonians~(\ref{eq:Ham}) and (\ref{eq:H_threequbit})
are still far from experimental realization. The closest analogy between (\ref{eq:Ham}) and an actual 
experimental qubit system is probably a chain of 
coupled flux qubits, which can be described by the Hamiltonian~\cite{maje05,dica09}
\be
H_{\rm flux chain} = \sum_{i=1}^N \Delta_i \sigma_x^i +  \sum_{i=1}^{N} K_{i,i+1} \sigma_z^i
\sigma_z^{i+1} + \sum_{i=1}^N \eps_i \sigma_z^i.
\label{eq:fluxchain}
\ee
Here $\eps_i$ denotes the magnetic energy, $\Delta_i$ the tunnel coupling energy of individual qubits and 
$K_{i,i+1} \sim M_{i,i+1} I_p^i I_p^{i+1}$ with $M_{i,i+1}$ the mutual inductance between the persistent 
currents $I_p^i$ and $I_p^{i+1}$ of qubits $i$ and $i+1$, respectively. 
In practice, local (in situ) tuning of the parameters $\Delta_i$, $K_{i,i+1}$ 
and $\eps_i$, as required by adiabatic evolution of Eq.~(\ref{eq:Ham}), is challenging, but promising 
progress is being made~\cite{paau09,dica09}. This gives hope for achieving control over individual qubits in chain 
and array-like qubit geometries and thereby brings experimental simulation of adiabatic quantum algorithms 
such as Exact Cover 3 closer.

\section{Conclusion}
\label{sec-conc}

In conclusion, we have calculated the energy level spectrum of the quantum Ising model in the presence of 
uniform beyond-nearest-neighbour interactions and an additional longitudinal field. We found that the gap 
between the ground state and the lowest-lying excited state increases with increasing number $M$ 
of neighbouring interactions (provided $M$ remains much smaller than the total number of qubits along the chain) 
and is approximately linearly proportional to $M$ for linearly decreasing interaction strength between the qubits 
along the chain. The increase of this gap with $M$, which persists in the presence of the additional weak longitudinal 
field, gives hope that the extended quantum Ising model is suitable for numerical - and in the future hopefully 
also experimental - simulation of quantum algorithms such as Exact Cover 3.

\acknowledgments

We thank J.H.H. Perk for useful comments. This work has been supported by the Netherlands Organisation for
Scientific Research (NWO). 

\appendix
\section{}
\label{appendix}

In this Appendix we investigate the validity of the perturbative approach that we used in Sec.~\ref{sec-long} 
and derive a criterion for the application of perturbation theory in terms of the number of qubits ($N$) and 
the number of nearest neighbours included ($M$). We first demonstrate that all odd-order corrections to 
the single-fermion energy levels are zero and then investigate even-order corrections. Our starting point 
is the third-order correction, which in general form is given by:
\bea
\nonumber  \delta E_{p}^{(3)} &=& \sum_{q \neq p} \sum_{r \neq p} \frac{\langle \theta_p | H_3 | 
\theta_q \rangle \langle \theta_q | H_3 | \theta_r \rangle \langle \theta_r | H_3 | \theta_p \rangle}
{\left( E_p - E_q \right) \left( E_p - E_r \right)} \\
&& - \langle \theta_p | H_3 | \theta_p \rangle \sum_{q \neq p} \frac{|\langle \theta_p | H_3 | \theta_q 
\rangle|^2}{\left( E_p - E_q \right)^2}.
\label{thirdcor}
\eea
It can be seen immediately that the second term is zero since it contains the matrix element $\la \theta_p 
| H_3 | \theta_p \ra$, 
which is zero (see Eq.~(\ref{firstcor})). The first term in Eq.~(\ref{thirdcor}) contains three separate matrix elements. 
Let us assume that in the first product $\theta_p$ is an even-fermion state, which is then coupled to odd-fermion 
states $\theta_q$ by $H_3$. This in turn restricts $\theta_r$ to the even fermion subspace. Applying the same 
reasoning to the last inner product we see that for this product to be non-zero $\theta_p$ should be an odd-fermion 
state. This is a contradiction with our starting assumption. Hence there is no combination of states that gives a non-zero 
outcome. This observation can be extended to all odd-order energy corrections and we conclude that these are 
therefore all zero~\cite{ovch03}.

We now investigate even-order corrections to the energy of the ground state and single-fermion states and 
use these to derive a criterion for the validity of perturbation theory. Our starting point is the second-order 
correction of the ground state, which can be found
by inserting Eqns.~(\ref{eq:Diagham_fullneighbours}) and (\ref{rk}) into Eq.~(\ref{02corr}). This yields:
\be
\delta E_0^{(2)} = - \frac{h^2}{2 N \Gamma}  \left[ \frac{1}{4 \Lambda_0} + \sum_{k=1}^{(N-1)/2} 
\left( \frac{1+\tan^2 \left( \frac{\pi k}{N} \right)}{\Lambda_k} \right) \right]. 
\label{dE0} 
\ee
After defining $\tilde{M} \equiv \sum_{j=1}^{M} j \lambda_j$, rewriting the sum using $m=N-2k$ and 
expanding around $m=0$ (which yields the largest contribution to the sum), Eq.~(\ref{dE0}) reduces to
\bea
\nonumber \delta E_0^{(2)} &\approx& \frac{h^2}{2 N \Gamma}  \sum_{\substack{
  m=1 \\
   m \; \textrm{odd}
  }}^{N-1} \frac{1+\left( \frac{2N}{m\pi} \right)^2}{\tilde{M} \left( \frac{m\pi}{N} \right)} \\ 
&\approx& -0.068\,  \frac{h^2 N^2}{\Gamma \tilde{M}}.
\eea
Analogously we obtain for the fourth-order correction:
\bea
\nonumber \delta E_0^{(4)} &=& \Gamma \sum_{k,l} \left( \frac{r_k}{2} \right)^2  \left( \frac{r_l}{2} \right)^2 
\left[ \Lambda_k^2 \left( \Lambda_k + \Lambda_l \right) \right]^{-1} \\
\nonumber &\approx& \frac{4}{\pi^5}\,  \frac{h^4 N^5}{(\Gamma \tilde{M})^3}  \sum_{\substack{
  m=1 \\
   m \; \textrm{odd}
  }}^{N-1}  \sum_{\substack{
  n=1 \\
   n \; \textrm{odd}
  }}^{N-1} \frac{m^{-4} n^{-2}}{m+n} \\
  &\approx& 0.0071\, \frac{h^4 N^5}{(\Gamma \tilde{M})^3}. 
\eea
We are interested in the energy gap between the ground state and the first excited states 
($N-2k \ll N$), which in the absence of the longitudinal field $h$ is given by
\be
\Delta_{eg,k} \equiv E_k^{(0)} - E_0^{(0)} = \Gamma \Lambda_k = \frac{\pi}{N} \Gamma \tilde{M} (N-2k),
\label{gap2}
\ee
and reduces to Eq.~(\ref{gap}) for the lowest-lying excited state (given by $k = (N-1)/2$). The second-order 
corrections to this gap between the ground state and 
the lowest-lying excited states ($N-2k \ll N$) are given by  (see Eq.~(\ref{eq:deltaDelta}))
\be
\delta E_k^{(2)} - \delta E_0^{(2)} = 2 \Gamma \frac{(r_k)^2}{\Lambda_k} = \frac{4}{\pi^3} \frac{h^2}{\Gamma} 
\frac{N^2}{\tilde{M}(N-2k)}.
\ee
For the general $(2n)^{th}$-order correction to the gap we find
\be
\delta E_k^{(2n)} - \delta E_0^{(2n)}  \propto \frac{h^{2n} N^{3n-1}}{(\Gamma \tilde{M})^{2n-1}}.
\label{Ek2n}
\ee
It follows from Eq.~(\ref{Ek2n}) that perturbation theory is valid for 
\be 
\frac{h}{\Gamma} \ll \frac{\tilde{M}}{N^{3/2}}.
\label{scaling}
\ee
Fig.~\ref{fig:mingap} in the main text shows a plot of $\Delta_{eg,min}$, which is directly proportional to $\tilde{M}$, 
as a function of the number of neighbouring interactions $M$. For exponential decay the value of $\tilde{M}$ converges 
for $M$ large (but still much smaller than $N$), but for linear coupling $\tilde{M}$ scales almost linear with $M$. 
For a given number of qubits $N$ and linear decay of interaction strength along the chain adding more nearest-neighbours 
thus enhances the range of validity of perturbation theory.


\begin{thebibliography}{99}
\bibitem{pfeu702}
This is the generalization of the situation encountered for $N$ even and $\lambda_1=1$ when only 
nearest neighbour coupling is included.~\cite{murg04,pfeu70}
\bibitem{murg2}
This has been noted before for the quantum Ising model with nearest-neighbour interactions only, 
see e.g. Ref.~\cite{murg04}.
\bibitem{bitstrings}
We verified that adding more instances only changes $p_E$ in the 4${\rm th}$ decimal (or smaller).
\end{thebibliography}
\end{document}